\documentclass[pra,twocolumn,showpacs,superscriptaddress,10pt]{revtex4-1} 
\usepackage{amsmath,amssymb,graphicx}

\usepackage{eqnarray}
\usepackage{hyperref}
\usepackage{fancyref}

\newcommand{\enqqote}[1]{``#1''}

\begin{document}
\title{Complete crossing of Fano resonances in an optical microcavity via nonlinear tuning}
\author{Martino Bernard}\email{Corresponding author: martino.bernard@unitn.it}
\affiliation{Centre for Materials and Microsystems, Fondazione Bruno Kessler, I-38123 Povo, Italy}
\affiliation{Department of Physics, Nanoscience Laboratory, University of Trento, I-38123 Povo, Italy}

\author{Fernando Ramiro Manzano}
\affiliation{Department of Physics, Nanoscience Laboratory, University of Trento, I-38123 Povo, Italy}

\author{Lorenzo Pavesi}
\affiliation{Department of Physics, Nanoscience Laboratory, University of Trento, I-38123 Povo, Italy}

\author{Georg Pucker}
\affiliation{Centre for Materials and Microsystems, Fondazione Bruno Kessler, I-38123 Povo, Italy}

\author{Iacopo Carusotto}
\affiliation{INO-CNR BEC Center and Department of Physics, University of Trento, I-38123 Povo, Italy}

\author{Mher Ghulinyan}
\affiliation{Centre for Materials and Microsystems, Fondazione Bruno Kessler, I-38123 Povo, Italy}

\begin{abstract}
We report on the modeling, simulation and experimental demonstration of complete mode crossings of Fano resonances within chip-integrated microresonators. The continuous reshaping of resonant lineshapes is achieved via nonlinear thermo-optical tuning when the cavity-coupled optical pump is partially absorbed by the material. The locally generated heat then produces a thermal field, which influences the spatially overlapping optical modes, allowing thus to alter the relative spectral separation of resonances. 
Furthermore, we exploit such tunability to probe continuously the coupling between different families of quasi-degenerate modes that exhibit asymmetric Fano-interactions. As a particular case, we demonstrate for the first time a complete disappearance of one of the modal features in the transmission spectrum as predicted by U. Fano [Phys. Rev. 124, 1866 (1961)]. 
The phenomenon is modeled as a third order non-linearity with a spatial distribution that depends on the stored optical field and the thermal diffusion within the resonator. The performed non-linear numerical simulations are in excellent agreement with the experimental results, which confirm the validity of the developed theory. 
\end{abstract}
\maketitle
\section{INTRODUCTION}
\label{sec:intro}
On chip microcavities have been proven of significant interest to the development of photonics for telecommunications. \cite{vahala_optical_2003,photonicsbook}.
Their ability to confine electromagnetic radiation and enhancement of light-matter interactions make them suitable for signal filtering and non-linear frequency generation processes, while the compatibility with the silicon platform opens the possibility for scalability \cite{sherwood2011scalable,asghari2011silicon,bogaerts2012silicon}. Delay lines, memories, modulators and frequency comb sources
have been demonstrated in microresonator devices  within the silicon platform \cite{matsko,liu2010ultra,xu_micrometre-scale_2005,foster_silicon-based_2011}.
For practical applications, in order to select desired wavelengths, a spectral tunability of the resonant features is required.
This is often achieved via electrical micro-heaters that exploit the thermo-optic effect to modify the refractive index of the environment, changing thus the optical path of the device \cite{dong2010low}. The thermo-optical tuning can be applied to a variety of materials systems, from silicon to silicon nitrides and oxynitrides, and is applicable to a wide range of operation wavelengths. This kind of approach, however, has a global action and does not allow to tune spectral channels separately.
Altering the transmission of a selected single resonance has been demonstrated, for example, by exploiting inverse Raman scattering in Si devices \cite{wen_all-optical_2011,wen_all-optical_2012}. As opposed to thermo-optical tuning, this effect is strongly selective to the operation frequencies, since the Raman frequency and bandwidth depend primarily on the material choice.

Here we propose an all-optical approach to detune relatively the resonances within a microresonator by exploiting the thermo-optical effect at a local level. 
The effect is demonstrated by applying it to a Whispering Gallery resonator -- waveguide system which exhibits Fano interference features \cite{Kivshar,matsko,lipson,20,22,oeHuang,praFano}. The continuous detuning is used to explore complete mode crossings between two resonances. As a particular example, this fine tuning approach allows to access the complete destructive interference point on one side of the Fano lineshape, where one of the resonances disappears from the spectrum, as predicted in \cite{praFano}.\\
The paper is structured as follows; Section \ref{sec:theo} starts with a brief theoretical summary of the system under study. The generic thermo-optic effect is described in \ref{sec:thermoptic} and then, in \ref{sec:nonlinear}, introduced into the dynamic equations for the field amplitude in the resonator to demonstrate the tuning of a single resonance. The model is then extended to interacting modes in \ref{sec:twofamilies}. In Sec. \ref{sec:expresults} we report the experimental results, while Sec. \ref{sec:simulations} exposes the numerical simulations that match the model with the experiment. Finally, Sec. \ref{sec:concl} gives the summary
of our results and the conclusions.

\begin{figure}[t]
	\centering
	\includegraphics[width=8.5cm]{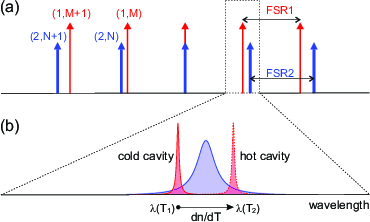}
	\caption{Schematic representation of the mode crossing possibilities: (a) The azimuthal modes of two radial families progressively shift at each increment of the azimuthal number due to the difference in FSR, possibly going through a crossing. (b) A continuous tuning of a doublet of resonances may be obtained via non-linearities, such as a localized thermo-optic effect. 
	}
	\label{fig:FSR}
\end{figure}
\section{Theory}
\label{sec:theo}
The wave equation inside a circular-shaped microresonator may be separated into three one-dimensional equations with quantum numbers $z, R$ and $M$ for the axial, radial and azimuthal coordinates, respectively. The resonator is considered to be single-mode in the axial direction ($z=1$).
The set of modes with different azimuthal number $M$ but with the same radial number $R=n$ is addressed as $n^{th}$-radial mode family, denoted with $R_n$. Here, we will consider only $R=1$ and $R=2$ for the radial direction. For each radial family, the frequency of the $M^{th}$ mode may be calculated as
\begin{equation}
	\omega_M=\frac{2\pi c M}{n_{eff}(z,R,M)L}~,
\label{eq:main}
\end{equation}
where $L$ is the geometrical path of the light in the resonator, typically the circumference, $n_{eff}(z,R,M)$ is the effective refractive index of the mode and $c$ is the light velocity in vacuum.

The resonator is excited and probed by a waveguide.
The transmission spectrum of the waveguide-resonator system exhibits series of negative peaks, due to the capture of the modes by the resonator, where the azimuthal modes are spaced by the Free Spectral Range (FSR)
\begin{equation}
	FSR(\omega_M)=\frac{2 \pi c}{n_g^M L},
\label{eq:FSR}
\end{equation}
with $n_{g}^M$ being the group velocity of the $M^{th}$ mode. Equation~\ref{eq:main} implies that, within a single radial family, modes with different $M$ lay at different frequencies, preventing photon exchange between such modes. 
When different radial families are present, however, there may be combinations of $R$ and $M$ which give spectrally overlapping modes (Fig.~\ref{fig:FSR}(a)).
While the spectral overlap may be an undesired feature~\cite{Lipson2}, it may also provide new interesting physics if an interaction among the degenerate modes is provided~\cite{praFano}.
 
Here, we will take into account the interference phenomena due to waveguide-mediated interactions among the modes of different radial families, leading to asymmetric Fano resonances and energy shifts of the modes. In order to observe the Fano interference, spectral overlap between a pair of modes belonging to different radial families is required. This means that $\Delta \omega=\omega_2-\omega_1$, relative distance between the central frequency of two modes $\omega_2$ and $\omega_1$ respectively, should be at most of the same order of the largest half-width at half maximum (HWHM) between the two modes,
\begin{equation}
	|\Delta \omega| \lesssim \max(HWHM_1,HWHM_2).
\label{eq:neighbours}
\end{equation}
Figure \ref{fig:FSR}(a) shows schematically that each radial mode family is a comb of peaks obeying Eq.~\ref{eq:main}, which are spaced spectrally by FSR according to Eq.~\ref{eq:FSR},
\begin{subequations}
	\begin{align}
	\omega_{M+l}^1 \simeq \omega_{M}^1+l\cdot FSR_1,\\
	\omega_{N+l}^2 \simeq \omega_{N}^2+l\cdot FSR_2,
	\label{eq:vernier}
	\end{align}
\end{subequations}
where $l=\pm 1, \pm2,...$
Since in general $n_{g}$ is different for different radial families, the FSRs will also differ. This means that the two radial mode combs will slide spectrally one against the other, similar to a Vernier scale \cite{griffel_vernier_2000}, passing though a crossing point. The relative detuning $\delta\omega^{12}$ of two modes in the couple is thus
\begin{equation}
	\delta\omega_l^{12}  = \omega_{M+l}^1-\omega_{N+l}^2 \simeq \delta\omega_0^{12}+l\cdot \Delta FSR_{12},
	\label{eq:detuning}
\end{equation}
where $\delta\omega_0^{12}=\omega_{M}^1-\omega_{N}^2$ is the relative detuning of $(MN)$-couple and $\Delta FSR_{12}=FSR_1-FSR_2$.

Having couples of modes of two radial families with different relative detuning permits to monitor the coupling between the two radial families when changing the $M$ with a discrete step of $\Delta FSR_{12}$ just by changing $M$. This has been studied in detail in our previous work ~\cite{praFano}.

The Fano-interaction of two resonances offers a rich physics and is extremely sensitive to both the degree and the sign of the detuning $\delta\omega^{12}_l$. The asymmetric Fano lineshape may vary abruptly with changing the detuning due to the fine interplay of the phases of individual resonances. For example, there exists a unique spectral point on one side of the composite resonance, where a particular destructive interference leads to a full suppression of one of the transmission features in the spectrum \cite{Fano1,praFano}. This situation may be observed experimentally in very fortuitous cases for which a given $\Delta FSR_{12}$ exactly satisfies the necessary condition. On the other hand, the particular point can be always observed if the detuning can be continuously varied within a single couple of resonances. 

In the following, we demonstrate a continuous way to detune the relative frequency of a couple of modes, allowing a complete mapping of the interaction within a single couple of resonances. Moreover, this approach permits us to tune finely and capture the particular point of resonance suppression.

\subsection{Thermo-optic effect} 
\label{sec:thermoptic}
\begin{figure}
	\begin{center}
		\begin{tabular}{c}
			\includegraphics[width=8.5cm]{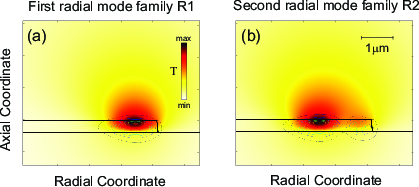} 
		\end{tabular}
	\end{center}
	\caption[]
	{ \label{fig:thermolocal}
		Simulated thermal distribution generated by  (a) the first and (b) the second optical radial family modes. The contour lines shows the modes electric field profiles. }
\end{figure}

Here we propose to use the thermo-optic effect~\cite{Baak69,Lipson2,FernandoThermobis}, as a tool to tune the relative frequency of the resonator's modes.
For small variations around room temperature, the thermo-optic coefficient $dn/dT$ may be defined as the first Taylor expansion of the refractive index of the material as a function of the temperature.
Heating the sample thus induces a variation in the effective refractive indices and coupling coefficients of the cavity modes, that are consequently shifted. 
Adding the thermo-optic coefficient $dn/dT$ into Eq.~\ref{eq:main} we obtain:
\begin{equation}
	\omega_M \approx \frac{2 \pi c M}{L n_{eff}}\left( 1-\frac{dn_{eff}}{dT} \frac{\Delta T}{n_{eff}}\right)=\omega_M^0+\delta \omega.
\label{eq:maindT}
\end{equation} 
With this definition, the usual positive thermo-optic coefficient makes the resonances shift toward the red (longer wavelengths) when heating the sample.
When a high-power laser (hereafter, pump, $p$) is used to excite a doublet of resonances, a part of the energy is absorbed by the material by heating locally the resonator exactly in the spatial region where the modes extend.
Heat then diffuses gradually over the whole device. 
Figure \ref{fig:thermolocal} shows a Finite Elements Method (FEM) simulation of the temperature distribution inside the resonator when the electrical field distribution function $\mathcal E^{p}(r)$ of the first and second radial family modes are used as the thermal source. The respective  spatial distribution functions of the electrical field of the modes $\mathcal E(r)$ are shown as contour lines in Fig.~\ref{fig:thermolocal}(a) and Fig.~\ref{fig:thermolocal}(b).

The thermal diffusion may be represented with a kernel functional $\chi(r,r')$ that takes into account for the absorption and thermal properties of the host materials. The temperature shift is also proportional to the square of the mode amplitude $\alpha^{p}$.
The thermal shift $\delta n(r)$ in the refractive index is therefore proportional to
\begin{multline}
\begin{split}
	\delta n(r) &\propto \int \chi(r,r') |\alpha^{p} \mathcal E^{p}(r')|^2 dr'\\
	&\propto |\alpha^{p}|^2 \int \chi(r,r') |\mathcal E^{p}(r')|^2 dr'.
\label{eq:dnr}
\end{split}
\end{multline} 
Alongside, the effect of $\delta n(r)$ on the spectral shift of the weakly excited probe mode (hereafter, called $s$) depends on the spatial overlap of the probe mode's electrical field distribution $\mathcal E^{s}(r)$ with the pump-generated  $\delta n(r)$ 
\begin{equation}
	\delta \omega \propto \int | \mathcal{E}^{s}(r)|^2 \delta n(r)  dr.
\label{eq:domega}
\end{equation}
This frequency shift may be rewritten in terms of a constant $g$ that incorporates the light-matter interaction and the spatial overlaps of the pump and probe modes mediated by the thermal kernel $\chi(r,r')$:
\begin{multline}
	\begin{split}
	\delta \omega 	&= |\alpha^{p}|^2 \int |\mathcal{E}^{s}(r)|^2 \int \chi(r,r')|\mathcal E^{p}(r')|^2 dr'~dr \\
					&= |\alpha^{p}|^2 g .
	\end{split}
	\label{eq:gint}
\end{multline}
A consequence of the proportionality of the spectral shift $\delta \omega$ to the mode intensity $|\mathcal{E}^{s}(r)|^2$ is that the effect is stronger for modes with an higher quality factor, which exhibit stronger field enhancement.

\subsection{Single family tuning}
\label{sec:nonlinear}
\begin{figure}[t]
	\centering
	\includegraphics[width=8.5cm]{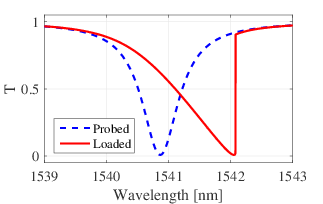}
	\caption{Resonant lineshape modification under a sweeping pump in the presence of optical non-linearity. The cold cavity spectrum (dashed line) is obtained with a weak probe. When sweeping the spectrum using a high power laser (solid line), the resonance shifts progressively due to the increasing non-linear effect, resulting in a spectrum with an apparent  discontinuity, where the cavity mode de-locks from the pump laser.}
	\label{fig:triangolo}
\end{figure}

As discussed in the beginning, for each radial family, the spectral position of unperturbed azimuthal modes is given by Eq.~\ref{eq:main}. The spectral shape and the waveguide-induced perturbation effect of a generic mode $i$ may be deduced plugging the unperturbed resonance frequency $\omega^0_i$ into the dynamic equation of the mode's intensity~\cite{praFano}:
\begin{multline}
\begin{split}
	\mathcal{L}_i(t)&\equiv i \frac{d\alpha_i}{dt}\\
	&=	\left[\omega_i^o + \Delta_{ii} - i \frac{\gamma_i^{\rm nr}+\Gamma^{\rm rad}_{ii}}{2}\right]\alpha_i+\bar{f}_i E_{\rm inc}(t)\\
	\label{eq:dy_eq_1}
\end{split}
\end{multline}
where $\alpha_i$ is the mode amplitude, $\Delta_{ii}$ is a self-interacting reactive term due to the presence of the coupling waveguide, $\gamma_i^{\rm nr}$ and $\Gamma^{\rm rad}_{ii}$ are the dissipative terms representing, respectively, the non-radiative and radiative losses of the mode, $\bar{f}_i$ is the coupling strength between the waveguide and resonator. Here, $E_{\rm inc}(t)$ is the incident field propagating in the waveguide, that we will take in the form  $E_{\rm inc}(t)=E_{\rm inc} e^{-i\omega_{\rm inc} t}$. The typical spectral shape of a resonance, calculated from Eq.~\ref{eq:dy_eq_1} in steady state and under weak excitation conditions, is a Lorentzian function (see, the cold cavity case in Fig.~\ref{fig:triangolo}).

Let us now introduce the intensity-dependent non-linearity $g$ of Eq.~\ref{eq:gint} in the resonator. The new dynamic equation for mode $i$ reads
\begin{equation}
	 i \frac{d\alpha_i}{dt}=\mathcal{L}_i(t)+|\alpha_i|^2 g \alpha_{i} .
	\label{eq:nl_eq_1}
\end{equation}
Consider now a dynamic pumping of the system, obtained for example by a strong tunable laser slowly scanning across the resonance from short to long wavelengths. When the pump reaches the resonance tail, the laser power enters the resonator enabling the last term in Eq:~\ref{eq:nl_eq_1}.
While the pump laser enters more inside the resonance peak, the non-linear term increases, and the resonance escapes from the laser line. The time-evolution of the intensity results therefore in the peculiar triangular shape ( Fig.~\ref{fig:triangolo}, full line). When the peak is surpassed, the non-linear term decreases, resulting in a sudden de-locking of the peak from the driving laser \cite{FernandoThermobis}. 
In addition to the self-interacting term $g$ introduced above, other terms $g_{ji}$ coupling the amplitude of $i^{th}$ mode to the intensity in the $j^{th}$ mode could be present, for example, when the system is pumped with a broadband source, which excites strongly more than one resonance at the same time.
In this case, the system evolution is described by a series of non-linear differential equations, coupled to each other by the non-linear terms $g_{ji}$:
\begin{equation}
  i \frac{d\alpha_i}{dt}=\mathcal{L}_i(t) + \sum\limits_{j} |\alpha_j|^2 g_{ji} \alpha_i.
\label{eq:FLT1}
\end{equation}

The non-linear interaction may be exploited to alter the spectral position of a resonance by acting on a different resonance of the same radial mode family. This can be achieved through a pump and probe experiment, where (a) a strong laser line excites a control single resonance, while (b) the linear spectral response of one or more resonances is monitored through a weak probe source. This last should be weak enough to guarantee that the terms which are  non-linear in the probe amplitude ($\sim |\alpha_i|^2$) are negligible.
In particular, it is possible to access the dynamics of a strongly pumped mode $M$ by following the spectral shift of an $(M+l)^{th}$ linear resonance. Indeed, for the non-pumped modes, the dynamic equation becomes parametric with respect to the field amplitude in mode $M$:
\begin{equation}
  i \frac{d\alpha_j}{dt}=\mathcal{L}_j(t) + |\alpha_M|^2 g_{Mj} \alpha_j.
\label{eq:FLT1_param}
\end{equation}
Since the spatial overlap integral of Eq.~\ref{eq:gint} is almost constant within the same radial family $R_1$, in Eq.~\ref{eq:FLT1_param} we may furthermore approximate all coefficients $g_{Mj}$ to be the same $g_{11}$ within a broad range of $M$'s. 

\subsection{Tuning of resonant doublets}
\label{sec:twofamilies}
Consider now two different radial mode families, $R_1$ and $R_2$.
Equation~\ref{eq:FLT1} is still valid, with the $i,j$ indices now running over all of the modes of both families. 
This time however, because of the different field distributions among the two radial families, more genuinely different coefficients will be present: the second radial family self-interaction term $g_{22}$ and a cross-family interaction term $g_{12}($=$g_{21})$.
The fact that the coupling constants are different for the two radial families gives rise to a net detuning when a single resonance is pumped.

In the following, we will consider doublets of resonances of two families in a region of spectral overlap, hence, the pump power will be split among two modes according to Eq.~(1) of~\cite{praFano}:
\begin{multline}
 \mathcal{F}^p_j(t)= i \frac{d\alpha^p_j}{dt}=
 \mathcal{L}^p_j(t)+\left(\Delta_{12} -  i \frac{\Gamma^{\rm rad}_{12}}{2}\right)\alpha^p_{3-j},
\label{eq:FL}
\end{multline}
where the index $j=1,2$ denotes all of the interacting modes of the doublet.
Joining Eq.~\ref{eq:FL} with Eq.~\ref{eq:FLT1} the doublets may be described with two coupled non-linear equations of motion for the pumped peaks, while the probe, acting on another doublet, may be described with a linear equation, parametric with respect to the pump fields. 
For the two pump ($p$) resonances it holds:
\begin{multline}
  i \frac{d\alpha^p_j}{dt}=
 \mathcal{F}^p_j(t) + \left( g_{jj} |\alpha^p_j|^2 +  g_{12}|\alpha^p_{3-j}|^2   \right) \alpha^p_j,
\label{eq:FLT2P}
\end{multline}
while for each couple of probe peaks ($s$), the probe equation reads:
\begin{multline}
  i \frac{d\alpha^s_j}{dt}=
 \mathcal{F}^s_j(t)+\left( g_{jj} |\alpha^p_j|^2 +  g_{12}|\alpha^p_{3-j}|^2   \right) \alpha^s_j.
\label{eq:FLT2S}
\end{multline}
The set of equations (\ref{eq:FLT2P}) and (\ref{eq:FLT2S}) provide a general description of the system during the pump and probe experiment. 
\section{Experimental results}
\label{sec:expresults}
\subsection{Sample description and experimental setup}
The theory described in the previous sections is supported by experimental results.
The samples studied in this work were realized using standard silicon microfabrication tools, as detailed in our previous works \cite{mherIEEE, verticup, praFano}.
The resonator is realized in silicon nitride material and is coupled vertically to the underlaying waveguide. The vertical coupling configuration plays a critical role, which is explained in the following. 

In a planar geometry, when all guiding components lay in the same plane, the waveguide has always a stronger coupling to the first radial family with respect to the second one. 
On the contrary, the vertical coupling technique permits to control the relative vertical alignment of the components. In particular, by placing the waveguide deeper under the resonator it is possible to invert the coupling trend between the $R_1$ and $R_2$ families such that the $R_2$ modes are coupled to the waveguide stronger than $R_1$ modes \cite{verticup, praFano}. The devices studied here were fabricated such that the waveguide is situated
400~nm under the resonator (vertical gap), 500~nm inward from the outer rim of the resonator (negative horizontal gap).

\begin{figure}[t]
	\centering
	\includegraphics[width=0.45\textwidth]{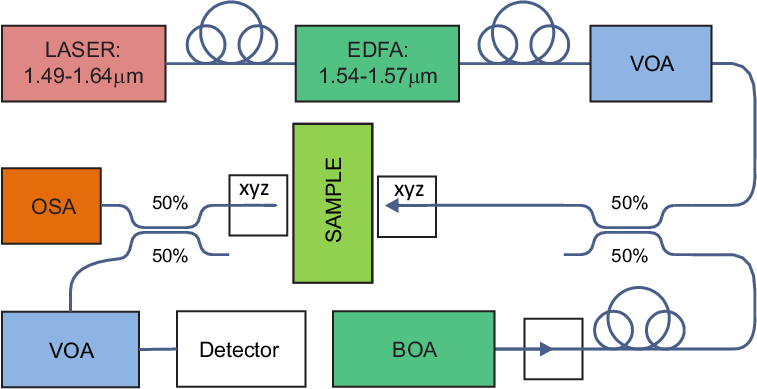}
	\caption[]
	{ \label{fig:ExpSetup}
		Experimental setup. A Tunable laser amplified with an Erbium Doped fiber Amplifier is mixed with the broadband signal of a BOA and shined into the sample with a taper fiber. The output is also collected with a taper fiber, split and fed to an Optical Spectrum Analyzer and a broadband germanium detector.}
\end{figure}

\begin{figure}[t]
	\begin{center}
		\begin{tabular}{c}
			\includegraphics[width=8.5cm]{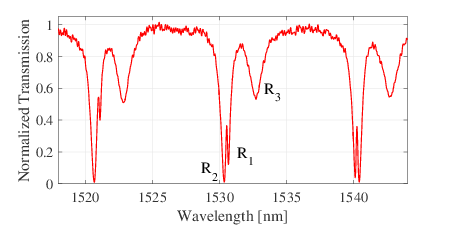}
		\end{tabular}
	\end{center}
	\caption[]
	{ \label{fig:aroundspectra}
		The experimental cold cavity spectrum of the resonator. Three azimuthal modes are present for the families $R_1$, $R_2$ and $R_3$. 
		The relative position of the $R_1$-$R_2$ doublet peaks transforms across the spectrum due to the difference in the respective FSR's. }
\end{figure}

The experimental setup, sketched in Fig.~\ref{fig:ExpSetup}, consists of a tunable laser (pump) and a broadband probe source (spontaneous emission from a booster amplifier, BOA). The laser is amplified through an Erbium Doped Fiber Amplifier (EDFA) and used as the high-power heating source. 
It is then mixed  with the broadband signal in a 3~dB coupler and injected into the waveguide through a lensed optical fiber. The transmitted signal at the output of the waveguide is then collected with another lensed fiber and 50:50 split between an Optical Spectrum Analyzer (OSA) to acquire the probe spectrum, and a germanium detector, that measures the overall transmitted intensity.
The detector measures the combined transmitted power of both the BOA and the pump, but the contribution of the second is several orders of magnitude larger than the first, permitting to measure the dynamic transmission spectrum of the pump. In addition, two Variable Optical Attenuators (VOA) are used to avoid damage on the detector and on the coupler fibers.
The OSA, on the other hand, allows the measurement of the probe transmission spectra due to the BOA broadband signal (Fig.~\ref{fig:aroundspectra}). In the following subsections we will describe the results of pump and probe experiments for two particular cases, namely, the full suppression of one of the doublet modes and the complete mode crossing in Fano resonances. In these experiments the sample was pumped with the EDFA in the region around a resonance doublet, while the broadband signal was used to probe the transmission of a different doublet.

\begin{figure}[b]
	\begin{center}
		\begin{tabular}{c}
			\includegraphics[width=8.5cm]{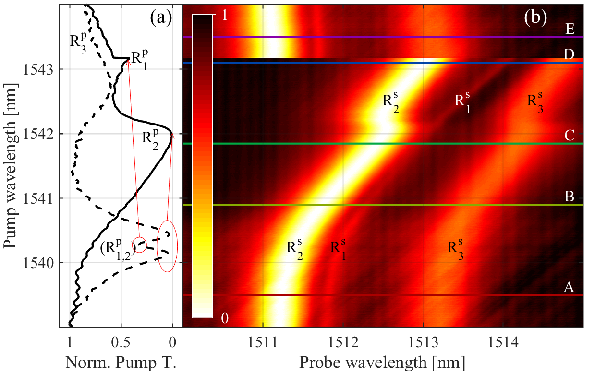}\\
			\includegraphics[width=8.5cm]{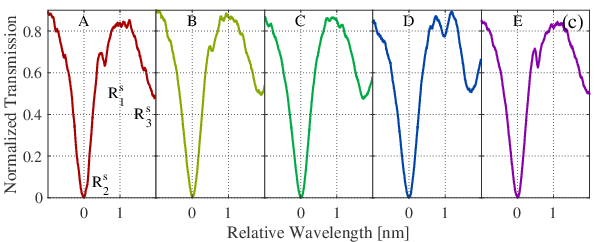}
		\end{tabular}
	\end{center}
	\caption[]
	{\label{fig:ExpMultipanel}
		Results of the pump and probe experiment. Panel (a) shows the cold (dashed) and hot (solid) cavity transmission spectra of the device around the strongly pumped resonance doublet. The thermo-optic non-linearity, induced by the pumped doublet, affects also the other resonances (b), allowing for a relative detuning of the peaks as shown by the transmission color map. (c) Selected transmission spectra show the transformation of the Fano resonance in the vicinity of the critical phase point, where a complete disappearance of the $R_1^s$ peak feature takes place (panel C).}
\end{figure}

\subsection{Resonant peak suppression}
\label{sec:nopeak}
Here we describe in detail the experiment where the rightmost doublet of Fig.\ref{fig:aroundspectra} is pumped, while a doublet with shorter wavelength is used as a probe. The chosen probe doublet forms a Fano resonance for which the relative detuning  $\delta\omega_0^{12}$ of two modes is very close to the critical interaction point, at which the narrow resonance is expected to be suppressed completely in the spectrum. 

Figure~\ref{fig:ExpMultipanel}(a) shows the dynamical transmission spectra of the resonances around 1540~nm as the 2~W pump laser is shined through, scanning from short to long wavelengths. The cold cavity spectrum (dashed line) is the same as in Fig. \ref{fig:aroundspectra}, while it's worth to note that in the pumped regime (solid line) both the peaks $R_2^p$ and $R_1^p$ get dragged by the thermal detuning, giving the typical triangular shape.
 
An interesting feature, indicating that interaction is present, is the smooth de-locking of the $R_2^p$ mode in favor of the $R_1^p$, right after the 1542~nm mark.
The de-locking of the $R_1^p$ mode, on the other side, is instantaneous, making a sharp jump partly eclipsed by the non-interacting third family. 
The effect is even more clearly visible in Fig.~\ref{fig:ExpMultipanel}(b), where the probe resonance spectra as a function of the pump laser wavelength is mapped. As the pump is swept from short to long wavelengths, the probe resonances are continuously detuned accordingly. Because of the difference in the overlap with the thermal fields generated from the $R_1^p$ and $R_2^p$ modes, the two probe peaks gain a relative spectral shift. This shift is gradually changing the original value of $\delta\omega_0^{12}$, bringing the system to the critical point at which a complete disappearance of the $R_{1}^s$ in the transmission spectra should take place. In fact, as a result of the fine tuning, we were able to validate experimentally the theoretical predictions from Ref.~\cite{praFano}.

The relative shifts and the mode-suppression are highlighted in Fig.~\ref{fig:ExpMultipanel}(c), where selected spectral cross sections A-E of Fig.~\ref{fig:ExpMultipanel}(b) are reported.
In panel A we show the cold cavity situation: the  sharp peak of $R_{1}^s$ family is at longer wavelength than the broad, almost critically coupled, $R_{2}^s$ resonance. When the pump laser starts sweeping across the $R_2^p$ resonance, the intra-cavity pump intensity increases progressively, which in turn affects the probe doublet. Due to this the probe mode $R_{2}^s$ initially detunes faster than $R_1^s$, so that their relative detuning decreases (panel B). 

Near the $R_1^p$ switching point (panel C) the interference between the two resonances becomes critical, and the resonant feature $R_1^s$ disappears completely. The detuning in the Fano interaction may thus be exploited as a mean to suppress one of the modes, permitting single mode operation in the range ~\cite{LipsonResControl}.

When the pump surpasses 1542~nm, de-locking from $R_2^p$, the power balance between the two pumped modes changes rapidly. As a consequence, now the $R_1^p$ mode starts to heat more, pushing the $R_1^s$ further away in the red side (panel D). Finally, when also the second resonance is de-locked, both pump and probe spectra switch back to the initial unloaded condition (panel E).
The whole experiment was repeated at different pump powers and on different couples of modes with analogous results.

\begin{figure}[t]
	\begin{center}
		\begin{tabular}{c}
			\includegraphics[width=8.5cm]{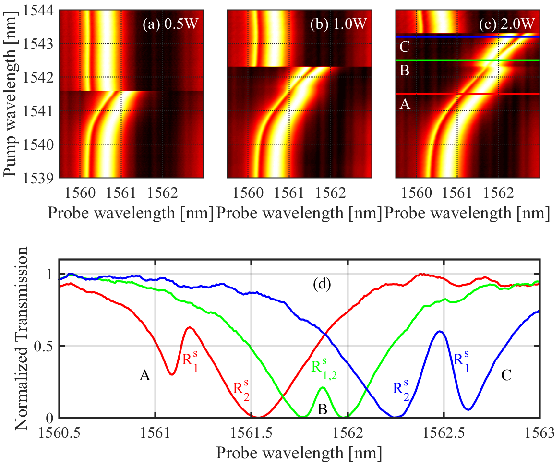} 
		\end{tabular}
	\end{center}
	\caption[]
	{ \label{fig:power}
		Pump and probe experiments demonstrating a complete crossing of the modes. Panels (a),(b) and (c) represent the same experiment under  different input power conditions 0.5~W, 1~W and 2~W, respectively. (d) The selected spectra, under 2~W pump, demonstrate three cases of the relative detuning, which changes from positive (A) to negative (C) passing through the $\delta\omega_0^{12}=0$ condition (B).}
\end{figure}

\subsection{Complete mode crossing}
\label{sec:cross}
In Fig.~\ref{fig:power} a complete crossing of a resonance doublet is reported, showing the behavior for different input powers. Note that the pump peaks of Fig.~\ref{fig:ExpMultipanel} and Fig.~\ref{fig:power} are the same, while a different couple of probe peaks is analyzed. In the latter case, the cold cavity spectrum has the $R_1^s$ peak on the blue side with respect to the wider $R_2^s$ resonance ($\delta\omega_0^{12}=\omega_0^{1}-\omega_0^{2}>0$). The figure reports the same experiment for different input powers.

At an input power of 0.5~W the two resonances of the probe doublet start red-shifting, and a progressive decrease in the relative detuning can be appreciated (Fig.~\ref{fig:power}(a)). However, this input power is not enough to drive the probe spectrum into the $\delta\omega_0^{12}\leq0$ condition. This happens because the pumped doublet de-locks from the laser, and the spectrum jumps back to the cold cavity situation. 

\begin{figure*}[]
	\begin{center}
		\begin{tabular}{c }
			\includegraphics[width=15cm]{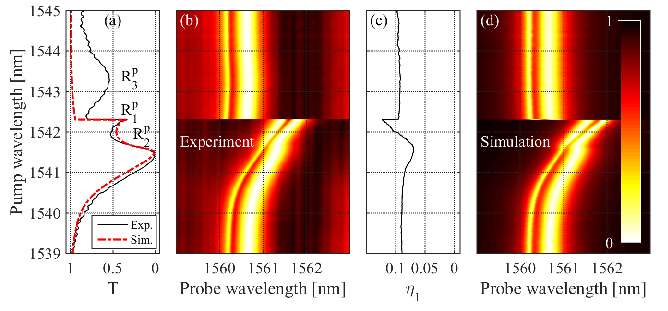}   \end{tabular}
	\end{center}
	\caption[]
	{ \label{fig:simmap}
		Experimental and simulated data of the pump and probe experiment. 
		(a)  the experimental pump transmission spectrum of the loaded cavity (black line) is simulated (dashed red) by inserting the cold cavity fit parameters into Eq.~\ref{eq:FLT2P}.
		In (b) the experimental transmission map as a function of both pump and probe wavelength is shown.
		In (c) we plot the relative coupling $\eta_1$ among the two radial family modes to the waveguide, extracted from results in panel (b). Successively, $\eta_1$ is used to compute the $\Gamma$ and $\Delta$ matrices of Eq~\ref{eq:dy_eq_1} to take into account also the changes in the coupling induced by the thermally-induced $\delta n(r)$.
		Panel (d) shows the transmission map as a function of both pump and probe wavelength using the simulated in (a) pump excitation for  Eq.~\ref{eq:FLT2S}.}
\end{figure*}

When the experiment is repeated for a 1.0~W of pump power (Fig.~\ref{fig:power}(b)), this time the pumped modes are dragged further in wavelength (late de-locking). This allows to reach the $\delta\omega_0^{12}=0$ situation for the probe doublet, which manifests as an EIT-like probe spectrum. 

Finally, we observed the complete mode crossing under a 2~W pump power (Fig.~\ref{fig:power}(c)). In this case, the relative detuning of probe modes not only vanishes but also changes the sign ($\delta\omega_0^{12}<0$). In Fig.~\ref{fig:power}(d) we plot selected spectra from Fig.~\ref{fig:power}(c) corresponding to three different cases of detuning $\delta\omega_0^{12}$: (A) the asymmetric Fano interference spectrum corresponds to the situation when the $R_1^s$ resonance is on the blue side of $R_2^s$ ($\delta\omega_0^{12}>0$), (B) the zero-detuning situation results into an EIT-like symmetric spectrum, when the aligned $R_1^s$ and $R_2^s$ modes interfere destructively, (C) $R_1^s$ finally surpasses $R_2^s$, giving rise to a Fano interference with sign-inverted relative detuning, $\delta\omega_0^{12}<0$.  

\section{Simulations}
\label{sec:simulations}
In this section we describe the numerical simulations, which we have performed in order to validate the observed experimental results. Making use of a numerical solver, the solutions of Eq.\ref{eq:FLT2P} and Eq.\ref{eq:FLT2S} are investigated. At first, the $\Delta_{ij}$ and $\Gamma_{ij}$ matrices are estimated together with the parameters $\gamma_j^{nr}$ and $\omega_j^o$ by fitting the cold cavity spectra of pumped doublet of resonances (the rightmost couple in Fig.~\ref{fig:aroundspectra}) .
The $g_{ij}^p$ matrix is then estimated up to an overall factor according to Sec.~\ref{sec:thermoptic}, making use of the FEM simulations shown in Fig.~\ref{fig:thermolocal}.
The $\Delta_{ij}$ and $\Gamma_{ij}$ matrices have been obtained using the same procedure as in \cite{praFano}. The $\Delta_{ij}$ and $\Gamma_{ij}$ matrices are calculated as $\bar\Delta \eta_{ij}$ and $\bar\Gamma \eta_{ij}$ respectively, where $\eta_{ij}=\eta_i\eta_j$ is the coupling weight matrix and $\eta_1,2$ are the relative coupling weights coefficients with the waveguide such that $\eta_1^2+\eta_2^2=1$.
The values of $\eta_i$, $\bar\Delta$ and $\bar\Gamma$ for the pump resonances of Fig.~\ref{fig:simmap} are reported in table~\ref{tab:simpar}.

\begin{table}[]
\centering
\setlength{\extrarowheight}{2pt}
	\begin{tabular}{|cc||cc|cc|}
	\hline
					&		&$R_1^p$&$R_2^p$&$R_1^s$&$R_2^s$\\
	\hline
	$\eta_i$		&		& 0.0723& 0.9974&0.0976$^*$&0.9952$^*$\\
	$\gamma_j^{nr}$	&[GHz]	& 4.400	& 20.581& 5.558	& 28.01	\\
	$\omega_j^o$	&[THz]	&194.532&194.374& 192.15& 191.92\\
	$g_{ii}$		&		& 1.00 	& 0.92	& 1.00	& 0.92	\\
	$\bar\Delta$	&[GHz]	&\multicolumn{2}{c|}{186.33}&
		\multicolumn{2}{c|}{193.44}\\
	$\bar\Gamma$	&[GHz]	&\multicolumn{2}{c|}{24.749}&
		\multicolumn{2}{c|}{18.567}\\
	$g_{12}$		&		&\multicolumn{2}{c|}{0.90}&
		\multicolumn{2}{c|}{0.90}\\
	\hline
	\end{tabular}
	\caption{Fit parameters of the cold cavity spectra of both pump and probe resonances, used in the simulations. In the probe case, the reported starred ($^*$) values of $\eta_{i}$ are the initial cold cavity values, but their evolution under the effect of the thermal field, reported in Fig.~\ref{fig:simmap} (c), was used in the simulation.}
	\label{tab:simpar}
\end{table}

In a next step, the non-linear pump equation, Eq.~\ref{eq:FLT2P}, is solved numerically to find the intensity distribution that generates the heat, in steady-state conditions, assuming the pump cold cavity spectrum parameters reported in Tab.~\ref{tab:simpar} as constant during the evolution. The solution is obtained moving the laser from short to long wavelengths.

Figure~\ref{fig:simmap}(a) shows that a good agreement was achieved between the experimental (black lines) and simulated (red dashed line) transmission of the pump. In Fig.~\ref{fig:simmap}(b) we show the experimental evolution of the probe spectrum as a function of the pump wavelength of panel (a). The experimental map is the same as is Fig.~\ref{fig:power}(b) and is reported here in order to facilitate the comparison with the simulated map.

The dynamic evolution of the relative coupling between the waveguide and the two radial family modes, that changes due to the thermo-optically altered refractive index $\delta n(r)$, is plotted in Fig.~\ref{fig:simmap}(c).
We note that the Eqs.~\ref{eq:FLT2P} and ~\ref{eq:FLT2S} do not consider explicitly non-linear variations of $\eta_1,\eta_2$, albeit the coupling among the waveguide and the resonator is affected by thermal detuning. 
The evolution of the relative coupling $\eta_1^{s}$ is therefore obtained by fitting the experimental probe spectra. 
The probe equation, Eq.~\ref{eq:FLT2S}, is linear with respect to the parametric functions of pump intensities. Therefore, Eq.~\ref{eq:FLT2S} is linearly solved for each pump wavelength, producing a 2D map reported in Fig.~\ref{fig:simmap}(d). The very good accordance between the measured and calculated maps indicates that the developed theoretical model is capturing all the details of the physical phenomenon.
\section{Conclusions}
\label{sec:concl}
In this work we have reported a joint theoretical and  experimental study on the possibility to tune cavity resonances via optically-driven material non-linearities. We actuated the thermo-optical effect locally around the mode extension by means of strong optical pump field, while the cavity dynamics was monitored via weak probe signal. The effect of the thermal non-linearity on the position and shape of the resonances has been exploited to tune the asymmetry of Fano resonances, realizing a complete mode crossing within a single doublet of modes. Moreover, we demonstrated the possibility to suppress one of these modes by fine tuning the system into a critical interaction condition. 

Theoretically the resonator-waveguide system has been modeled via a set of non-linear equations that take into account the local heating. The developed model was validated by numerical simulations which are in excellent agreement with experimental results. Further improvements to the model are expected taking into account the coupling variations induced by the thermo-optic effect also in the pump equations and possible non-linear losses.

Our theoretical approach can be generalized and extended to other material nonlinearities and photonic systems. For example, it may be possible to exploit ultrafast Kerr nonlinearities in order to achieve high-speed modulation of the spectral response in multimode resonator devices. 

\section*{ACKNOWLEDGEMENTS}
We acknowledge financial support from the Autonomous Province of Trento, partly under the Call {\em "Grandi Progetti 2012"}, project {\em "On silicon chip quantum optics for quantum computing and secure communications - SiQuro"}.\\



\end{document}